\documentclass[conference]{IEEEtran}

\usepackage{cmap}
\usepackage[utf8]{inputenc}
\usepackage[english]{babel}

\usepackage{indentfirst}
\usepackage{amsmath,amssymb,amscd,amsthm}
\usepackage{mathtools}
\usepackage{multirow}
\usepackage{multicol}
\usepackage{hyperref}
\usepackage{graphicx}
\usepackage{caption}
\usepackage{subcaption}
\usepackage{siunitx}
\usepackage{tikz}
\usetikzlibrary{shapes,shapes.geometric,arrows,fit,calc,positioning,automata}
\usetikzlibrary{arrows}
\usetikzlibrary{shapes.multipart}
\usetikzlibrary{decorations.pathreplacing}
\usetikzlibrary{patterns}
\usetikzlibrary{arrows.meta}

\usepackage{cite}

\begin{document}
\IEEEoverridecommandlockouts
\newcommand\copyrighttext{%
\footnotesize \textcopyright \enspace 2018 IEEE. Personal use of this material is permitted. Permission from IEEE must be obtained for all other uses, in any current or future media, including reprinting/republishing this material for advertising or promotional purposes, creating new collective works, for resale or redistribution to servers or lists, or reuse of any copyrighted component of this work in other works. DOI: \href{https://doi.org/10.1109/EnT-MIPT.2018.00014}{10.1109/EnT-MIPT.2018.00014}
}
\newcommand\copyrightnotice{%
\begin{tikzpicture}[remember picture,overlay]
\node[anchor=south] at (current page.south) {\fbox{\parbox{\dimexpr\textwidth-\fboxsep-\fboxrule\relax}{\copyrighttext}}};
\end{tikzpicture}%
}

\title{Clock Drift Impact on Target Wake Time \\ in IEEE 802.11ax/ah Networks
	\thanks{The research was supported by RFBR grant 18-07-01356 a.}
}

\author{
	\IEEEauthorblockN{Dmitry Bankov\IEEEauthorrefmark{1}\IEEEauthorrefmark{2}, Evgeny Khorov\IEEEauthorrefmark{1}\IEEEauthorrefmark{2}, Andrey Lyakhov\IEEEauthorrefmark{1}\IEEEauthorrefmark{2} and Ekaterina Stepanova\IEEEauthorrefmark{1}\IEEEauthorrefmark{2}}
	\IEEEauthorblockA{\IEEEauthorrefmark{1}Institute for Information Transmission Problems, Russian Academy of Sciences, Moscow, Russia\\
		\IEEEauthorrefmark{2}Moscow Institute of Physics and Technology (State University), Moscow, Russia\\
		Email: \{bankov, khorov, lyakhov, stepanova\}@iitp.ru}
}

\maketitle
\copyrightnotice
\begin{abstract}
In the Internet of Things scenarios, it is crucially important to provide low energy consumption of client devices.
To address this challenge, new Wi-Fi standards introduce the Target Wake Time (TWT) mechanism.
With TWT, devices transmit their data according to a schedule and move to the doze state afterwards.
The main problem of this mechanism is the clock drift phenomenon, because of which the devices cease to strictly comply with the schedule.
As a result, they can miss the scheduled transmission time, which increases active time and thus power consumption.
The paper investigates uplink transmission with two different TWT operation modes.
With the first mode, a sensor transmits a packet to the access point (AP) after waking up, using the random channel access.
With the second mode, the AP polls stations and they can transmit a packet only after receiving a trigger frame from the AP.
For both modes, the paper studies how the average transmission time, the packet loss rate and the average energy consumption depend on the different TWT parameters.
It is shown that when configured to guarantee the given packet loss rate, the first mode provides lower transmission time, while the second mode provides lower energy consumption.
\end{abstract}


\section{Introduction}
\label{sec:introduction}

Currently, the Internet of Things (IoT) is a popular concept, which aims at providing reliable connectivity to swarms of cheap and power-efficient devices.
To enable massive IoT, several technologies have been developed. One of them is described in the IEEE 802.11ah amendment to the Wi-Fi standard.
This amendment presents the Target Wake Time (TWT) mechanism, which enables frame transmissions between two stations (STAs) at scheduled intervals and switching them to the doze state in the remaining time. Since in the doze state the radio is switched off, power consumption is negligible. The apparent price for this is disabled transmission and reception of any signal. 
The IEEE 802.11ah market currently faces some problems. However, being very promising,  TWT has been adopted by the IEEE 802.11ax amendment \cite{11ax}, which now is being actively developed and is going to replace IEEE 802.11ac as the flagman Wi-Fi standard.
For that reason, we investigate the TWT efficiency within IEEE 802.11ax.

In the paper, we consider a scenario with an access point (AP) and several stations (STAs) willing to exchange frames.
The essence of the TWT is that the STAs can agree in advance on schedule.  Thus they can separate their transmissions, avoid collisions and reduce extra power consumption caused by transmission retries and long carrier sense during alien frames. Moreover, transmissions of power-limited sensors can be separated from transmissions of common devices with heavy traffic. For many applications, sensors transmissions are very rare. For example, the AP can wait for some meter report every day at 9:00 (further $T_{target}$). Since both the sensor's and the AP's clocks are not ideal and scheduled transmissions can intersect, which leads to collisions and degrades performance. Specifically, the Wi-Fi standard allows the clock drift not exceeding 100 ppm, i.e., 100 us per 1s. So, the maximum relative deviation from the assigned wake-up time can reach $ \delta T = T * 200 \ ppm $, where $T$ is the wake-up period of a sensor device. 
So the main problem of the TWT mechanism is the clock drift effect.

Taking into account clock drifting, to guarantee the absence of collisions, one should reserve the channel for $\delta T$ before the scheduled TWT and $ \delta T + T_{data}$ after it, where $T_{data}$ is the expected duration of the data frame.
For example, such a channel reservation can be done with the CTS-to-self mechanism for short time intervals, or using the Restricted Access Window \cite{khorov2018two} functionality in IEEE 802.11ah networks, while in IEEE 802.11ax it can be done by making all the devices transmit only by OFDMA \cite{bankov2018ofdma} and not scheduling them transmissions in the protected interval.

Regardless of the mechanism used to reserve the channel, if T = 1 hour, then $\delta T = 3600 * 200 * 10^{-6} = 0.72$ s which is much higher than the average transmission time in a Wi-Fi network.
Reserving the channel for such a long time to transmit only one frame significantly decreases the network performance, because most of the protected time the channel remains idle.

If the protected interval length is reduced, the sensor transmissions can intersect, which potentially causes collisions. Moreover, if outside the protected interval the traffic is saturated, then the collision probability becomes significantly high.

To solve this problem, one can allocate scheduled transmissions side by side, making the guard interval between them shorter than needed for a single transmission. Thus, even if two transmissions overlap, they likely resolve the collision during the next transmission attempt. Since no other devices transmit the awake time will be small. So the power consumption increase is not so dramatic as in the case of collisions with heavy traffic. 

At the same time, the guard interval should not be too small, to allow STAs to resolve collisions quickly without forming a collision avalanche.
Therefore, there is a problem of the TWT parameter selection in order to maximize the network performance, because it is necessary to assign sensors' transmissions not very far apart and not very close to each other.
To solve this problem, we study the dependency of the average transmission time, the packet loss rate and the average energy consumption on different TWT parameters.

The rest of the paper is organized as follows.
Section \ref{sec:mechanism} briefly describes the TWT mechanism and the method of random access to the channel.
In Section \ref{sec:scenario}, we define the considered scenario and formulate the problem statement.
Section \ref{sec:papers} briefly reviews the related papers.
We run simulation and study the numerical results on the TWT efficiency in Section~\ref{sec:numerical}.
Section \ref{sec:conclusion} concludes the paper.

\section{Channel access and power saving in Wi-Fi networks}
\label{sec:mechanism}

\subsection{Channel access}
Wi-Fi devices use Enhanced Distributed Channel Access (EDCA) to transmit frames, including the TWT ones. 
Briefly, EDCA works as follows.
When a frame enters an empty queue, the STA senses the channel and, if it is idle, the STA transmits the frame.
If the channel is busy, the STA randomly draws a backoff, an integer number from the range from $0$ to $CW_r $, where $CW_r$ is the contention window and $r$ is the retry counter.
Initially, $r$ is zero and $CW_0$  equals  $CW_{min}$ which is 15 by default. 
The value of the backoff counter is decremented by one whenever the channel has been idle for the time slot $\Delta$.
When the channel is busy, the backoff counter is suspended.
If the channel is idle for the Arbitration Interframe Space (AIFS), the STA resumes the backoff countdown.
When the backoff counter becomes zero, the STA transmits the frame.
The STA that received a frame shall send an acknowledgment (ACK) to the sender Short Interframe Space (SIFS) after receiving.
If the transmitter receives the ACK, it considers the frame to be successfully delivered and proceeds to the next frame, if any.
If no ACK arrives during $AckTimeout$ after sending the frame, the STA considers that the frame is lost.
In this case, the STA increases the retry counter and selects a new value of backoff counter from a new window defined as following:
\[
CW_r =
\begin{cases}
CW_{min}, & r = 0,\\
\min \left\{2 (CW_{r - 1} + 1) - 1, CW_{max} \right\}, & r > 0,
\end{cases}
\]
where $CW_{max}$ is the maximal value of the contention window (by default, equal to 1023).
If the retry counter reaches the retry limit $RL$ (by default equal to 7), the frame is dropped from the queue.

The more STAs simultaneously try to transmit their data, the higher is the collision probability and the time required to resolve the collisions.
The TWT mechanism allows a STA to regulate the number of STAs simultaneously accessing the channel, and thus to increase the probability of data delivery and to reduce the energy and channel resources consumption of the sensors.

\subsection{Target Wake Time}
The TWT mechanism lets a STA and an AP agree in advance on the time instant (namely, the Target Wake Time, TWT), when the STA will be awake, and the TWT Service Period (TWT SP), which is the time interval after TWT, during which the AP and the STA can exchange frames.
Now let us describe TWT in more detail.
In order to configure the TWT mechanism, the AP and the STA exchange special frames.
These frames contain information about the agreement or refusal to establish the TWT with specific values of the TWT  parameters mentioned above.  Apart from that, the list of parameters includes the Trigger flag, which determines the mode in which frame exchange sequence starts. 

\paragraph{Polling mode (PM)}
If the Trigger flag is set to one, the AP polls the STAs at the scheduled time.
Specifically, before the scheduled time, the sensor  STA wakes up and waits for a trigger frame from the AP during the TWT SP.
SIFS after receiving the trigger frame, the sensor sends data. After the data are delivered, the sensor can switch to the doze state, if no more frames need to be transmitted.

\paragraph{Non-polling mode (NPM)}
If the Trigger flag is set to zero, the STAs wake up at the scheduled time, transmit their data frames with EDCA and switch to the doze state afterward.

\section{Network Scenario and Problem Statement}
\label{sec:scenario}

\begin{figure}[!b]
	\center{\includegraphics[width=0.8\linewidth]{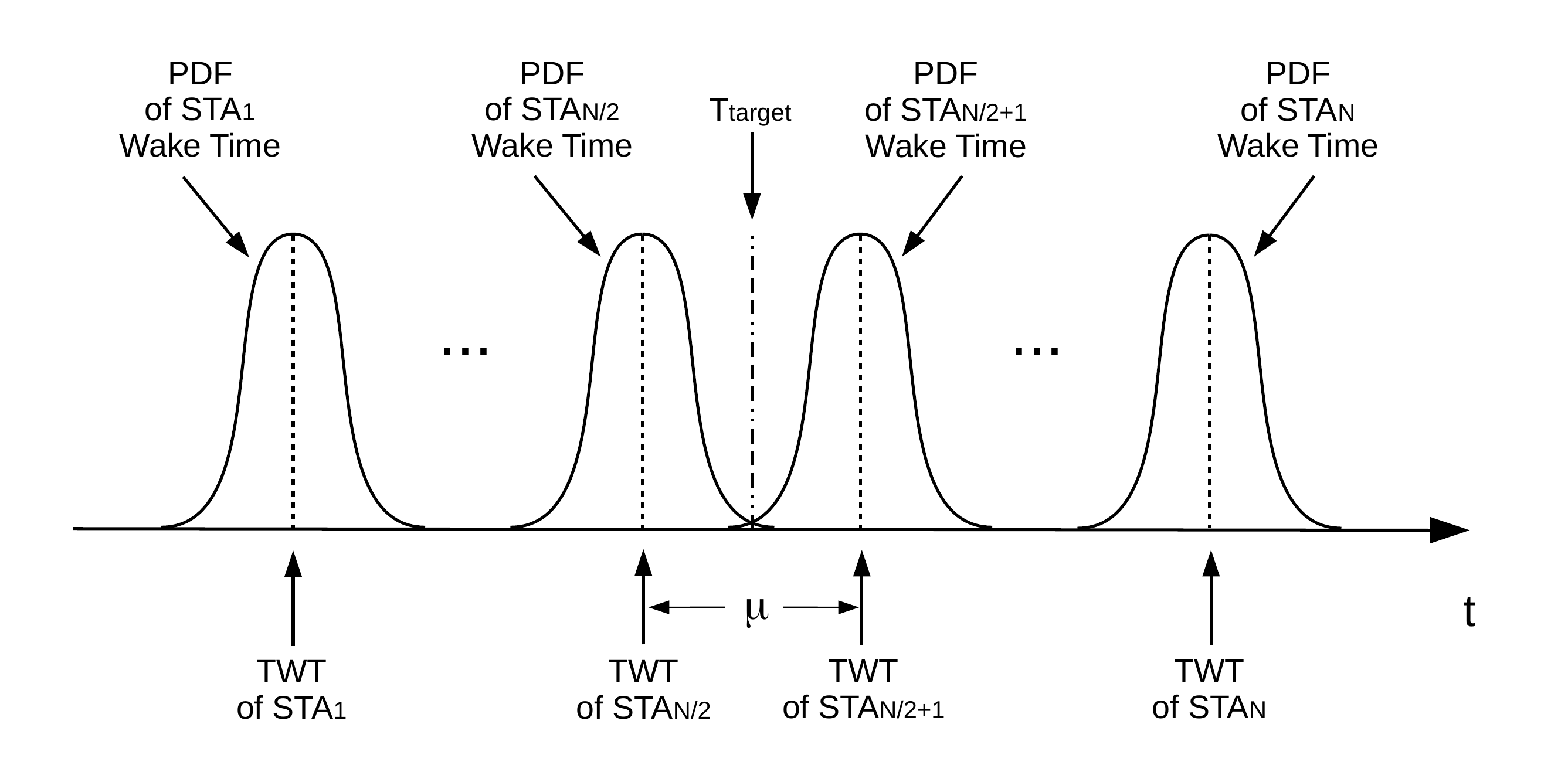}}
	\caption{Target and real wake time.}
	\label{ris:waketime}
\end{figure}

Consider a Wi-Fi network consisting of an AP and N sensor STAs, which use the TWT mechanism to transmit information to the AP.
Let the STAs complete association and the TWT establishment.
At the scheduled TWTs, each STA has one data frame, and all data frames are of the same size.
As discussed in Section \ref{sec:introduction}, the AP schedules their TWTs around a target time $T_{target}$, but separates them by equal time intervals in order to decrease the contention for the channel.
Let the interval between successive TWTs of different STAs be equal to $\mu$, and the STAs' clock drift from the assigned TWT be normally distributed with the variance $\sigma^2$, as shown in Fig. \ref{ris:waketime}.

In this paper, we study both operation modes: PM and NPM. Apart from that, we consider two cases: when all the STAs can hear each other, and when some STAs are hidden from the others. Also,  we may take into account the capture effect.
With the capture effect, if the AP detects a preamble of a frame while receiving a less powerful frame, it can discard the current frame and begin receiving the new one.
This effect can reduce the number of collisions, since if the AP switches to receiving a more powerful transmission, then one of the frames can be successfully received, but if there is no switching, both the collided frames are lost.
The described cases are different from each other, so the strategy optimal for one case may provide bad results for others. 

We formulate the following problem within these scenarios:
\emph{To investigate the dependency of the average transmission time, the probability of frame delivery and the average energy spent by each STA on the TWT settings, the clock drift distribution, parameters of the random channel access, the presence of capture effect and hidden STAs}.

\section{Related Papers}
\label{sec:papers}
Nowadays, there are some papers \cite{khorov2015survey, park2015ieee, afaqui2016ieee} which touch the TWT operation. However, there is no published research on the performance evaluation of the TWT mechanism.
The authors of \cite{tian2016implementation} write about their plans to implement the TWT mechanism on the ns-3 simulation platform \cite{ns-3}.
Paper \cite{tian2016evaluation} mentions that the TWT mechanism can be used to reduce the number of collisions, but does not describe how to use the TWT for this purpose.

There is a large number of studies on the ways to estimate the clock drift and to synchronize the clocks, as in \cite{romer2001time}.
The study \cite{song2008wirelesshart} suggests that during each frame exchange with the sensors, the AP should evaluate (and send to them in the ACK) the value of the current time deflection for clock adjustment.
If the initiator of the frame exchange is the AP, then it should send the current value of its clock in the data frame.
This method is inapplicable in the scenario considered in this paper since for successful transmission it is necessary to know in advance the deviation of the clock.
The authors of \cite{tjoa2004clock} present an algorithm that estimates the speed of the clock drift.
For that, the sensors should track and accumulate the time deviation between the scheduled and real transmissions for each of their neighbors.

\section{Analyses of the Numerical Results}
\label{sec:numerical}

\begin{figure}[tb]
	\center{\includegraphics[width=190pt,height=155pt]{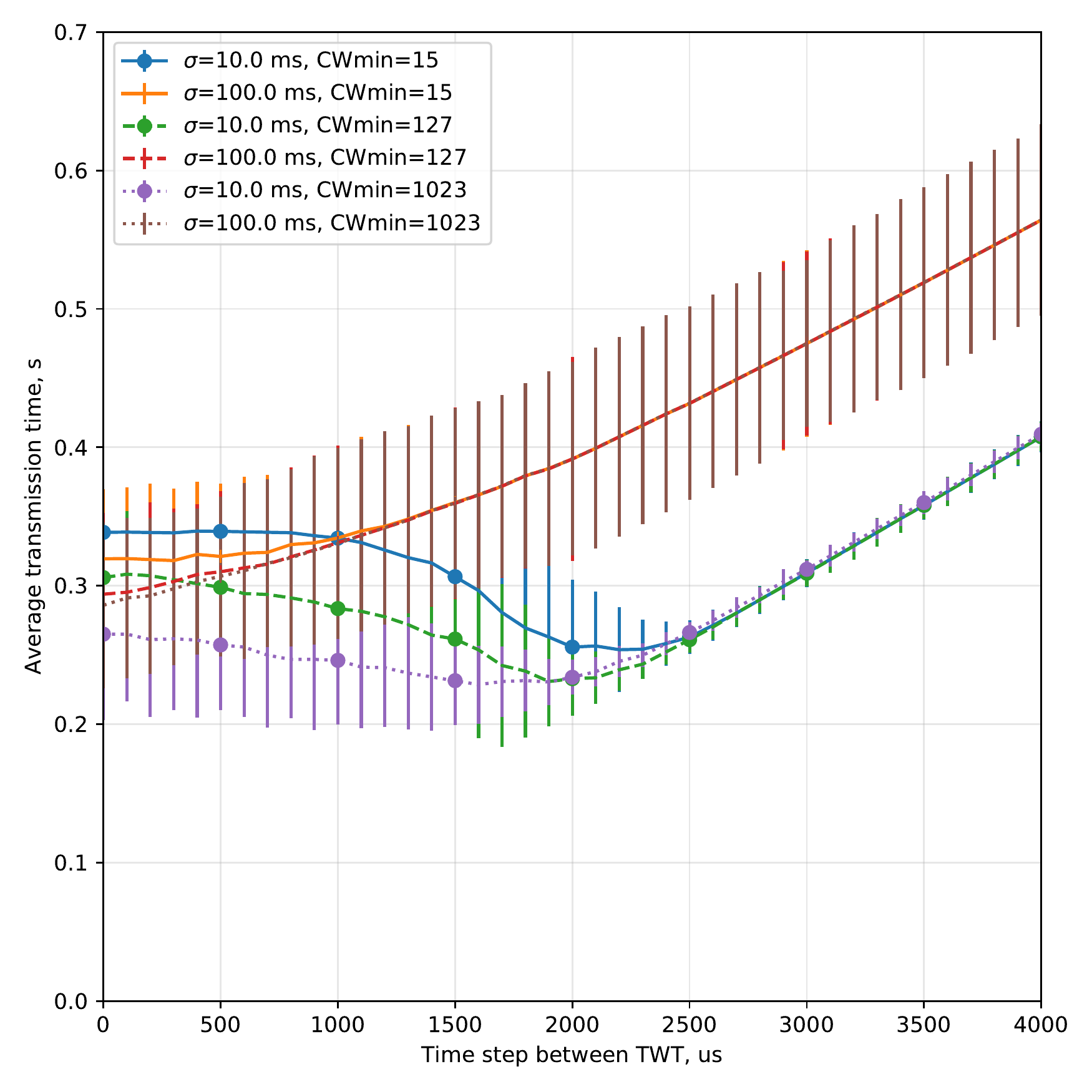}}
	\caption{Dependency of the average transmission time on time step $\mu$ between TWT for the NPM without hidden STAs and without capture effect.}
	\label{ris:endtime}
\end{figure}

To evaluate the effectiveness of TWT, we use the scenario described in the Section \ref{sec:scenario}. For that, we have implemented in the ns-3 simulation platform \cite{ns-3} the trigger frame and the state machine for PM. Note that if a STA wakes up after the AP sends the trigger frame to it, the STA cannot transmit data in the current TWT SP, because it needs to receive the trigger frame to send data.
Therefore, the AP should assign TWT SPs to STAs in advance, so that sensors have time to wake up and the trigger frame is received even taking into account the clock drift effect.
Therefore, with PM, the AP assigns TWTs several $\sigma$ before sending the trigger frame.
Further, we refer to this parameter as ``awake offset''.

In our simulation, we use the most reliable modulation and coding scheme MCS0, which results in the duration of the data transmission of 1480 us.
The TWT SP is 1 s.
The carrier sense range for STAs is 65 m. All STAs are placed uniformly around the AP within a circle with a radius of 5 m in the ``scenario without hidden STAs'' or 50 m in the ``scenario with hidden STAs''.
The carrier sense range for STAs is 65 m, so some STAs are hidden for others when the radius is 50 m.

Fig. \ref{ris:endtime} shows the dependency of the average transmission time by a group of N STAs on the time step $\mu$ between their TWTs for the NPM.
By the average transmission time, we denote the time from the first scheduled TWT to the end of the last successful transmission of the last STA.
If the step between TWTs is small, many stations simultaneously try to transmit their frames, which increases the number of collisions and the average transmission time.
A high value of $CW_{min}$ reduces contention and the average transmission time because of the decrease in collision probability. For small steps between consequent TWTs, high values of $\sigma$ provide the same effect, since they spread transmission attempts over time.
It should also be noted that if the TWT step $\mu$ is higher than two successive transmission lengths, all dependencies become linear and have the form $T = N * \mu + \sigma$.

As explained further, the linear part of the plot corresponds to the packet delivery ratio (PDR) close to one, therefore to minimize the used channel resources, and at the same time to guarantee packet delivery, one should set the minimum possible step between TWTs at which the dependency has a linear form.

For PM, we obtained similar results not shown because of paper space limitation. 
The main difference is that the linear part of the plot does not depend on $\sigma$, because the data transmission occurs strictly after receiving of the trigger frame, which is sent by the AP at the scheduled time.

Let us consider the dependency of PDR on the step between STAs' TWTs for the NPM.
When the step is less than doubled transmission duration, PDR can be far from 1. For example, at $CW_{min}=15$, $\sigma=10$ ms and the step less than transmission duration, PDR is below 90\%. It means that even retransmissions cannot avoid loses because of the collision avalanche effect. Apart from low PDR, this area also has much higher energy consumption.

\begin{figure}[tb]
	\center{\includegraphics[width=190pt,height=155pt]{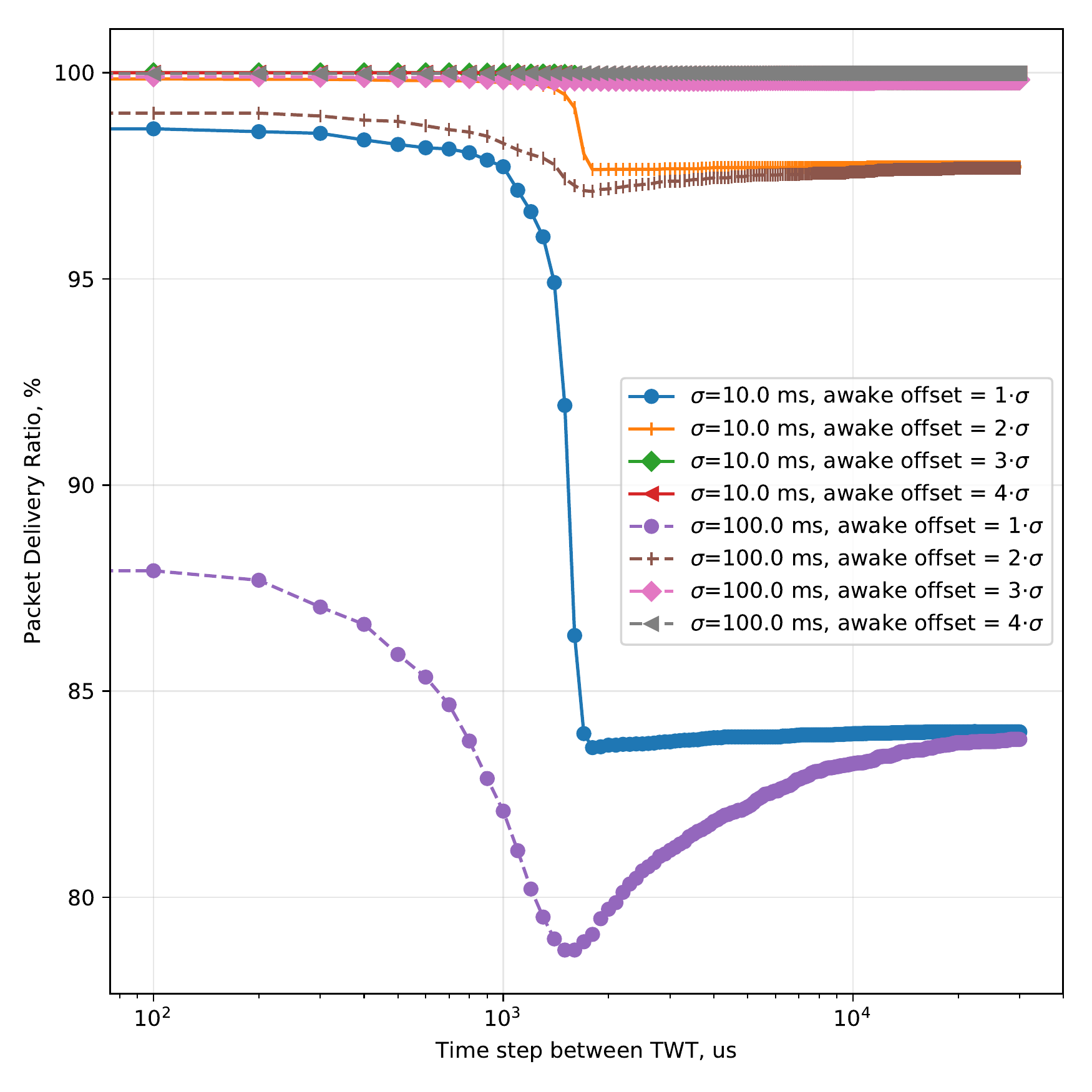}}
	\caption{Dependency of the PDR on time step $\mu$ between TWTs for the PM with hidden STAs and without capture effect.}
	\label{ris:trigger_plr}
\end{figure}

Fig. \ref{ris:trigger_plr} represents the same dependence for the PM, and it is strikingly different from the dependence for the NPM.
Accumulation of the trigger frames in the AP queue with  small steps between TWTs causes the trigger frame transmission to be delayed and the STAs successfully receive the trigger frames even if their clock drift makes them awake later than the designated TWT.
This leads to the significant increase of the PDR with small steps between TWTs comparing with the NPM.
When the step between TWTs increases, the trigger frame delay in the AP queue decreases, so the time by which the STA can be late with awakening to successfully accept the trigger frame and, hence, the PDR are reduced.
With further increase in the step between TWTs, the PDR tends to a horizontal line, the height of which depends on the distribution of sensor clock drifts and on the ``awake offset'' parameter, i.e.,the earlier the STA wakes up, the higher probability of the trigger frame reception and the data frame sending.

\begin{figure}[tb]
	\center{\includegraphics[width=190pt,height=155pt]{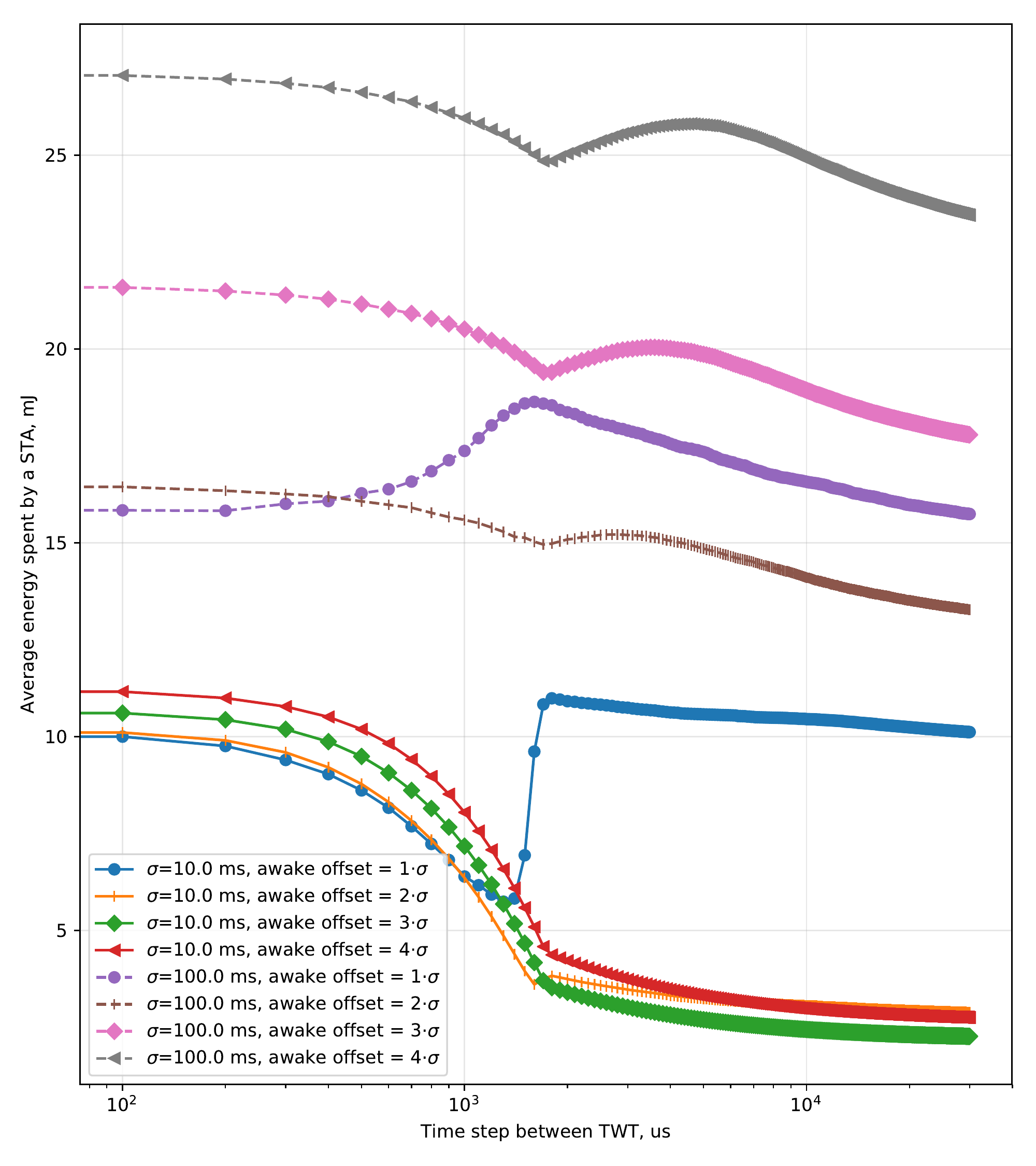}}
	\caption{Dependency of the average energy spent by a STA on time step $\mu$ between TWTs for the PM without hidden STAs and without capture effect.}
	\label{ris:trigger_energy}
\end{figure}

\begin{figure}[tb]
	\center{\includegraphics[width=190pt,height=155pt]{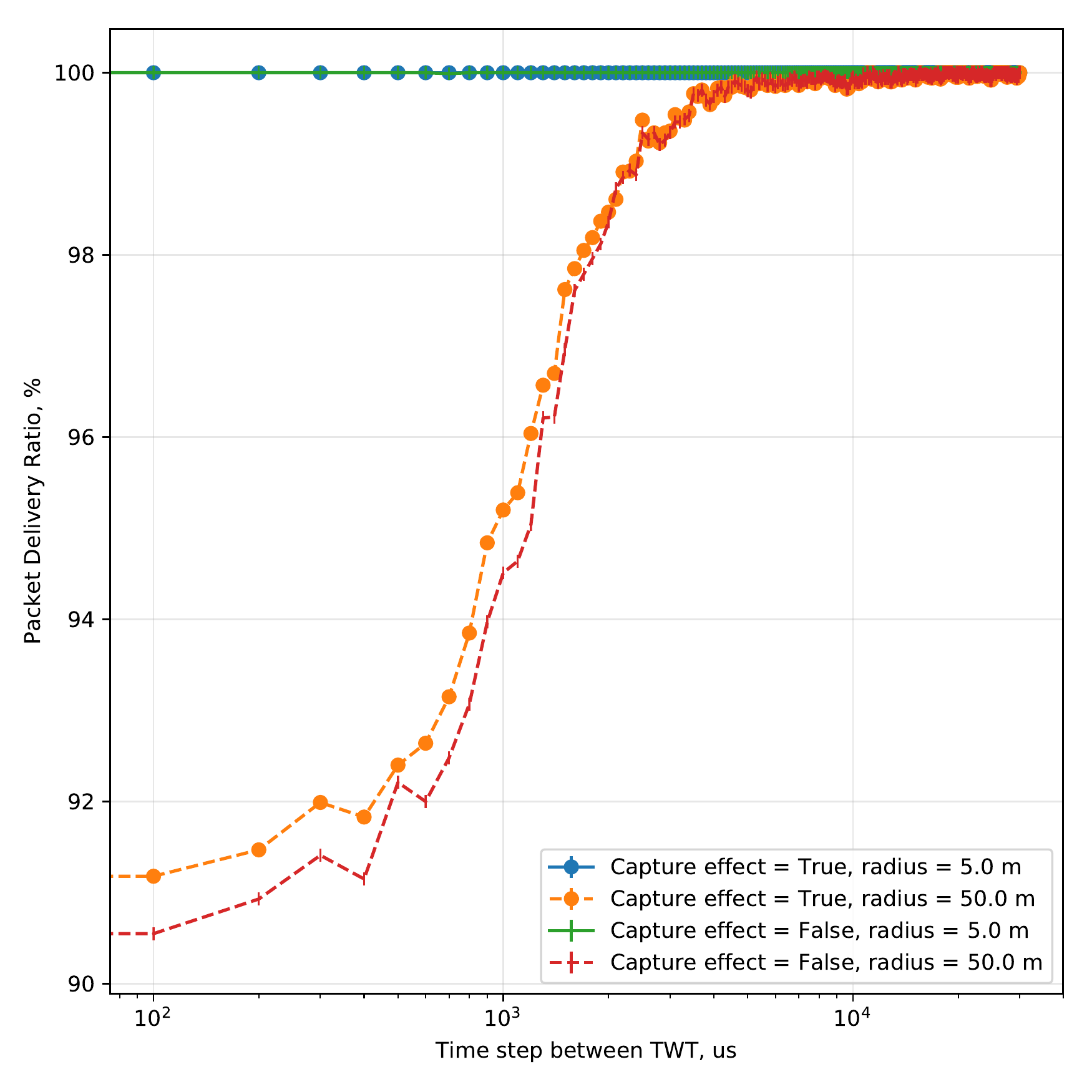}}
	\caption{Dependency of the PDR on time step $\mu$ between TWTs for the NPM with $\sigma = 100$ ms and $CW_{min} = 15$.}
	\label{ris:plr_all}
\end{figure}

Note that with PM, the energy consumption is higher than with NPM.
The reason is the unnecessary listening to the channel, during which a STA wastes much energy receiving frames not addressed to it.
We can point out three sources of such a listening for the PM.
The first one is when the STA wakes up after the AP sends it a trigger frame, e.g., when the awake offset is low.
In such a situation, the STA does not transmit its data frame, cannot switch to the doze state right after the transmission, and waits for the end of the TWT SP instead.
So with the PM, the TWT SP duration is an essential factor of the energy consumption.
The second source is when the STA wakes up too long before the trigger frame transmission, because of large awake offset.
However, the awake offset cannot be made too low, since it will lead to packet losses.
The third source is when the step between the TWTs is small, and the STA has to wait until the AP transmits all the trigger frames in its queue and receives the corresponding data frames before the trigger frame intended for the STA.
The NPM is free from these disadvantages, but when the step between TWTs is small, the contention for the channel access and the resulting frame retransmissions also increase the energy consumption.
However, the minimal energy consumption of NPM is lower than that of PM.

\begin{figure}[tb]
	\center{\includegraphics[width=190pt,height=155pt]{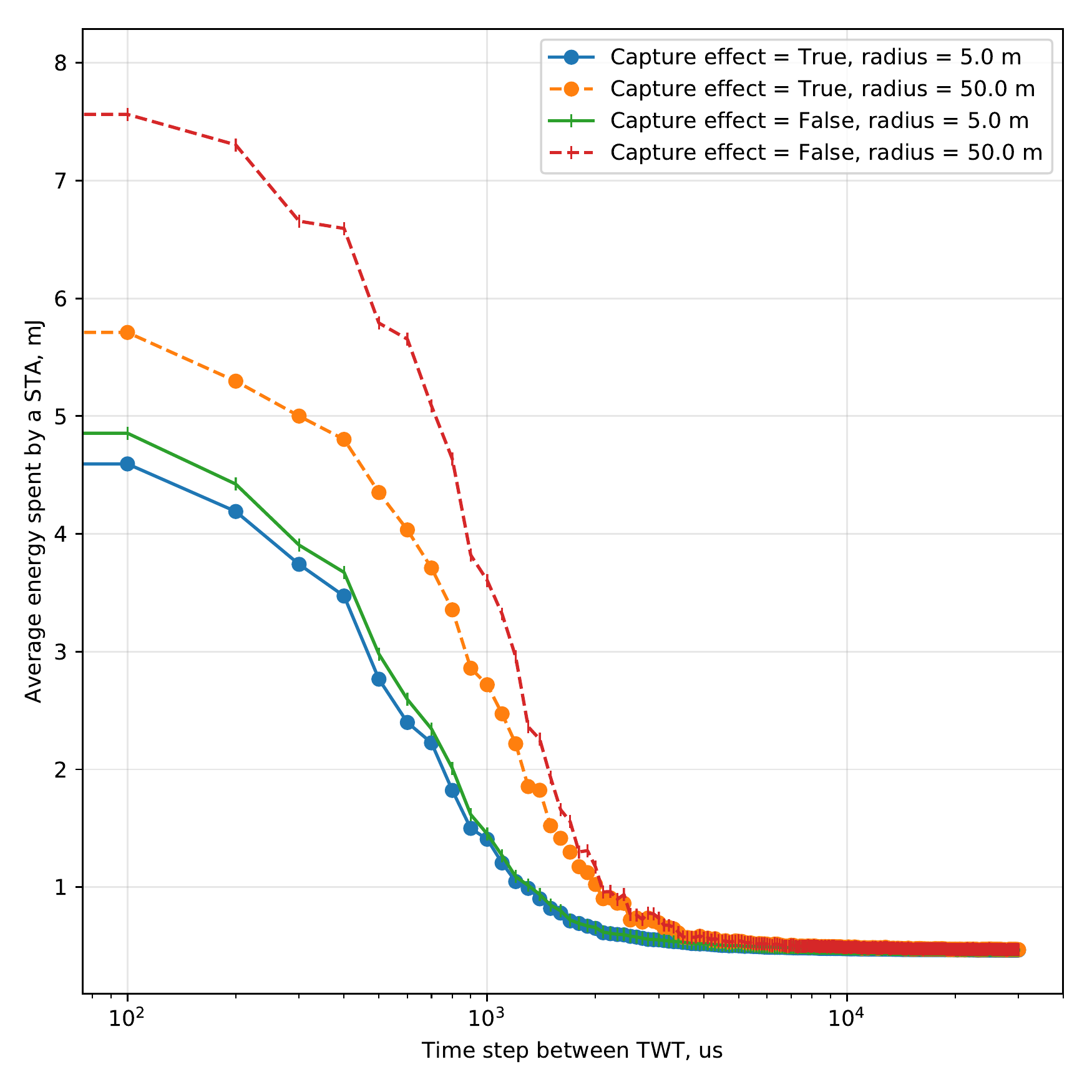}}
	\caption{Dependency of the average energy spent by a STA on time step $\mu$ between TWTs for the NPM with $\sigma = 100$ ms and $CW_{min} = 15$.}
	\label{ris:energy_all}
\end{figure}

Now let us compare the results obtained with hidden STAs and without them.
With the NPM, the hidden STAs increase the number of collisions as shown in Fig. \ref{ris:plr_all}, while with the PM there are not collisions.
Therefore, with the NPM, the energy consumption increases noticeably when hidden STAs appear in the network as shown in Fig. \ref{ris:energy_all}, while with the PM the results of the experiments remain unchanged.

Similar observations can be made for scenarios with the capture effect and without it.
The capture effect is noticeable in scenarios with a large number of collisions, reducing the channel resources consumption.
Such an effect is particularly evident in Fig. \ref{ris:energy_all}, which shows that for $CW_{min} = 15$ and small step between TWTs with the NPM, the capture effect reduces the energy consumption by 1.4 times in the case with hidden STAs.
If we consider the cases with a small number of collisions, then the capture effect does not give a visible improvement.
With the PM, collisions cannot occur because of the frame exchange sequence, so the capture effect does not affect the obtained results.

From all the plots presented above, we draw the following conclusion:
if the network performance indicator is the energy consumption of the sensors, the Trigger flag should be set to zero, $CW_{min}$ should be set as high as possible, and the step between  TWTs should be set to two successful frame transmissions;
if the network performance indicator are the PDR and the average transmission time by a group of STAs, the Trigger flag should be set to 1 and the AP should assign the TWT SPs to STAs not less then $3\cdot\sigma$ before the scheduled trigger frame transmission.

\section{Conclusion}
\label{sec:conclusion}

In this paper, we have studied the Target Wake Time (TWT) mechanism within the IEEE 802.11ax amendment of the Wi-Fi standard.
TWT allows an AP to coordinate the transmissions of STAs by scheduling them the time intervals when they should be awake and may transmit the data frames.

The main problem of this mechanism is the clock drift, which causes the STAs to digress from the schedule.
It leads to collisions, retries, the increased power consumption of the STAs and packet losses.
We have studied the packet delivery ratio, STAs energy consumption and the average data transmission time when the TWT is used for uplink data transmission in the presence of the clock drift for two operation modes.

The first mode is the polling mode when a STA can transmit its data frame only after receiving a trigger frame from the AP.
With this mode, the data frames are transmitted without collisions, and, as a result, it takes a short time to deliver data frames from a group of STAs to the AP.
However, the clock drift can make a STA wake up after the AP sends it the trigger frame and thus lead to the packet loss and high energy consumption because the STA has to wait for the end of the scheduled time interval instead of switching to the doze state right after the frame transmission.
To deal with this problem, we propose the STAs to wake up before their TWT, and thus to ensure that they do not miss the trigger frame.

The second mode is the non-polling mode when a STA can transmit its data as soon as its TWT comes.
With this mode, contention for the channel between the STAs is an essential factor of the network performance, so it is necessary to set the largest possible $CW_{min}$ and to set the time step between the consecutive TWTs of different STAs corresponding to the chosen network performance indicator:
to reduce the STAs' energy consumption it is necessary to set the step between their TWTs as large as possible, while to reduce the average transmission time spent by the group of STAs, the step between their TWTs should be as low as possible with the PDR achieving its maximum value.
As a particular trade-off between these two cases, it is suggested to set the step between TWTs equal to two successful data frame transmissions.

The comparison of the studied modes shows that when the TWT is configured in such a way to guarantee the PDR close to one, the first mode provides lower data transmission time, while the second mode results in lower energy consumption.

\bibliographystyle{IEEEtran}
\bibliography{biblio} 
\end{document}